\documentclass[]{article}

\usepackage[affil-it]{authblk}
\usepackage{a4wide}
\usepackage{amsmath}
\usepackage{multirow}
\usepackage{booktabs}
\usepackage[ruled,vlined]{algorithm2e}
\usepackage{array}
\usepackage{color}
\usepackage[normalem]{ulem}
\usepackage{hyperref}
\usepackage{amssymb}
\usepackage{graphicx}
\usepackage{amsfonts}
\usepackage[english]{babel}
\usepackage[sort&compress,square,comma,authoryear]{natbib}

\begin{document}
\bibliographystyle{apalike}

\title{Toward using GANs in astrophysical Monte-Carlo simulations}

\author[1]{Ahab~Isaac}
\author[1]{Wesley~Armour}
\author[1]{Karel~Ad\'{a}mek \thanks{E-mail address: \texttt{karel.adamek@eng.ox.ac.uk}} }
\affil[1]{Oxford e-Research Centre, Department of Engineering Sciences, University of Oxford, 7 Keble road, OX1 3QG, Oxford, United Kingdom}

\maketitle

\begin{abstract}
Accurate modelling of spectra produced by X-ray sources requires the use of Monte-Carlo simulations. These simulations need to evaluate physical processes, such as those occurring in accretion processes around compact objects by sampling a number of different probability distributions. This is computationally time-consuming and could be sped up if replaced by neural networks. We demonstrate, on an example of the Maxwell-J\"{u}ttner distribution that describes the speed of relativistic electrons, that the generative adversarial network (GAN) is capable of statistically replicating the distribution. The average value of the Kolmogorov-Smirnov test is 0.5 for samples generated by the neural network, showing that the generated distribution cannot be distinguished from the true distribution.
\end{abstract}

\section{Introduction}
The concept of Generative adversarial networks (GAN), where the neural network (generator) is not trained directly on the data but rather through a proxy network (discriminator), has been successfully used in generating samples from underlying, often multidimensional, probability distributions in many applications\citet{chakraborty2023years}. However, GANs are less often used to generate samples from probability distributions for scientific purposes. Previously, GANs were used in the generation of 1D probability distribution with varying success as demonstrated by \citet{zaheer2017gan}. Recently GANs have been used to simulate particle showers in electromagnetic calorimeters by \citet{PhysRevD.97.014021}.

When simulating astrophysics processes around compact objects, particularly simulation of accretion flows and thick accretion discs, Monte Carlo simulations are often used \citep{Schnittman_2013}. Among many different probability distributions that are required to describe the state of the simulated system is the Maxwell-J\"{u}ttner distribution (MJD), describing an electron's speed in high energy plasma. This distribution is used in evaluations of Compton scattering during radiative transfer, Bhabha scattering that describes electron-positron scattering, or in many other fields \citep{10.1063/1.4919383}. The Maxwell-J\"{u}ttner distribution is given by
\begin{equation}
    P_\mathrm{MJ}(\gamma)=\frac{\gamma^2\beta}{\Theta K_\mathrm{2}\left(1/\Theta\right)}\exp\left(-\frac{\gamma}{\Theta}\right)\,,
    \label{eqa:MJ}
\end{equation}
where $\gamma$ is relativistic Lorentz factor, $\beta=v/c$ is normalised velocity of an electron, and $K_\mathrm{2}$ is the modified Bessel function of the second kind. Lastly, $\Theta=k_\mathrm{B}T/(mc^2)$, where $T$ is the temperature of the gas, $k_\mathrm{B}$ is Boltzmann constant, $m$ is the electron mass, and $c$ is the speed of light. The parameter $\Theta$ represents the ratio of kinetic and rest energy of the electron and indicates how relativistic the electron speed distribution is. If MJ distribution~\ref{eqa:MJ} is evaluated in normalised speed $\beta$ it's support is limited to an interval $\left[0,1\right[$

We demonstrate that it is possible to use GANs to draw samples from the MJ distribution and that per sample of 100 elements statistical test we have used is not able to distinguish between true and generated samples. 

\section{Network and training}
We have used GAN to train a multi-layer perception network to generate a sample from MJ distribution. The GAN approach requires two networks to be trained at the same time, a generator and a discriminator. Each sample produced by the generator contains 100 numbers which are then classified by the discriminator to either be from true distribution or not. Both the generator and discriminator are learning at the same time. The loss function is calculated based on the discriminator's performance and used to train both networks.

For the generator, we have used two hidden layers each 400 neurons in size, an input and output layer of 100 elements. The input is a vector of uniformly distributed pseudo-random numbers. The ratio of 4:1 regarding the size of the hidden layers and input/output layers is one that produces the best results. A smaller number of neurons in the hidden layer does not produce correct results. Larger values, on the other hand, do not substantially improve results for high values of $\Theta$. The activation function used was $\tanh()$. Other activation functions (hardtanh, SELU, SiLU, LeakyReLU, ELU) did not lead to similar or better results.

For the discriminator, we have used two hidden layers, 40 and 20 neurons wide, with a binary output layer. The input layer is the size of the generator output, which is a vector of 100 elements. As an activation function, we have used LeakyReLU. 

The Maxwell-J\"{u}ttner distribution was normalised into the interval $\left[-1,1\right[$ so that it coincides with the support of $\tanh()$ activation function. 

For training, we have used the binary cross entropy as our loss function. For both the generator and discriminator we have used ADAM as an optimiser with learning rate $l_\mathrm{r}=0.002$. The batch size was 512 samples. Training steps were as follows: First real MJ distribution is sampled using the rejection method; next a vector of uniformly distributed pseudo-random numbers is generated using NumPy built-in generator (PCG64), and a fake distribution is produced by the generator; after this the discriminator classifies both true and fake samples and the loss function is calculated on both groups; the training of the discriminator and generator follows, and then cycle repeats. The discriminator is trained on a loss function calculated on both true and fake samples, whilst the generator is trained only on a loss function calculated on fake samples. There is one training step for both the generator and discriminator. For each value of $\Theta$, we have trained a separate network.

Training progressed until the loss function stabilised, and then the best-performing network (set of weights and biases) was selected using the KS test applied to data composed from multiple generated samples. The networks selected are presented in the result section.

\section{Results}
\begin{figure*}[htb]
    \centering
    \includegraphics[width=.45\linewidth]{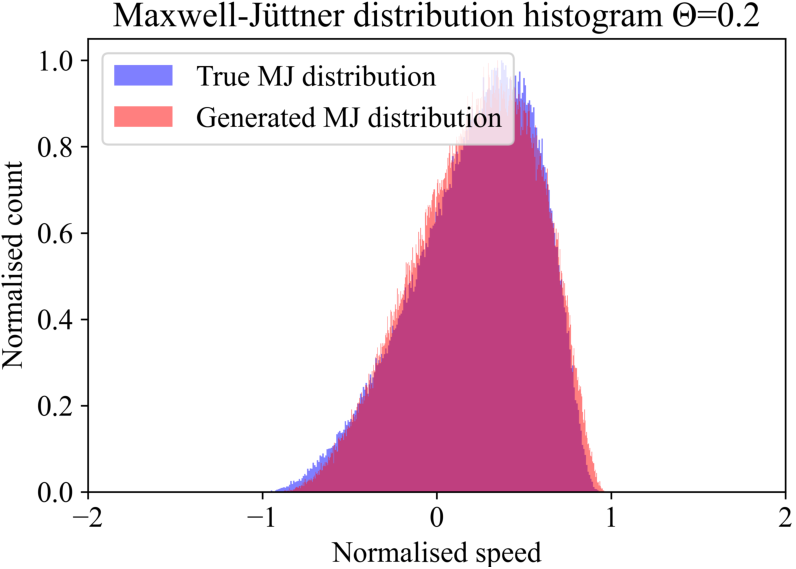}\hfill%
    \includegraphics[width=.45\linewidth]{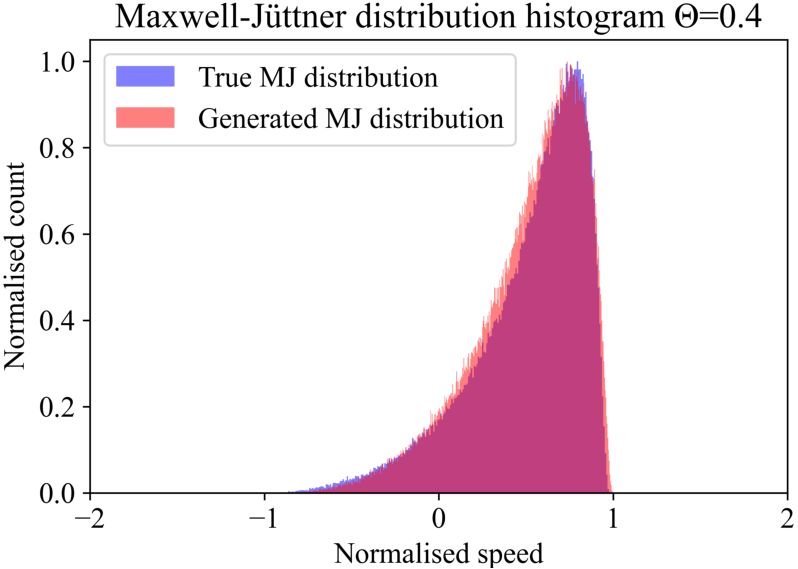}\\
    \includegraphics[width=.45\linewidth]{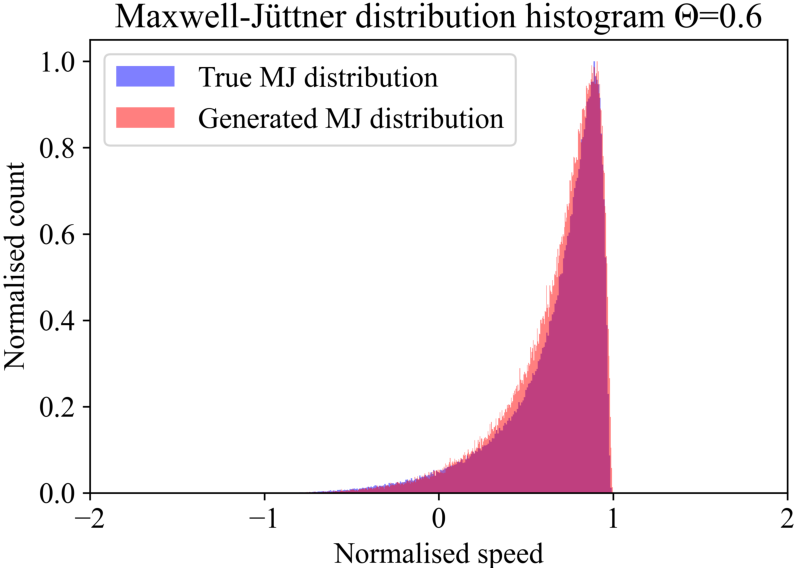}\hfill%
    \includegraphics[width=.45\linewidth]{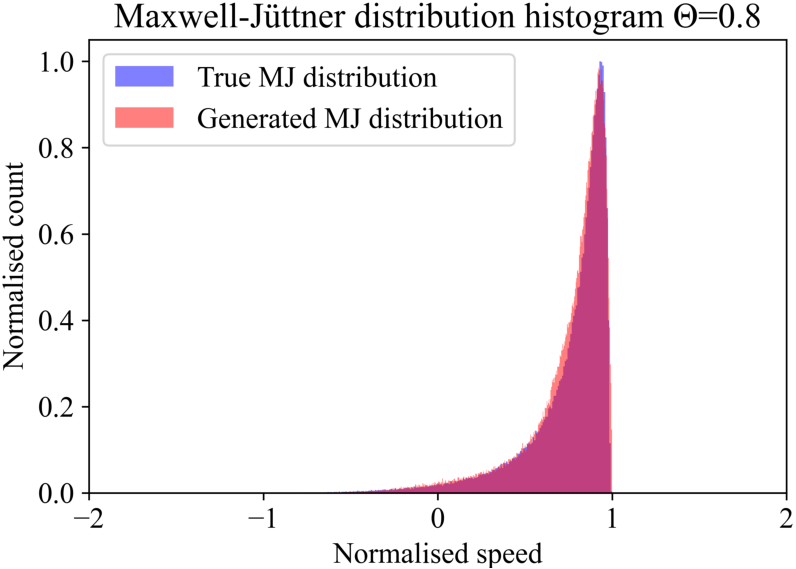}
    \caption{\label{fig:hst} Comparison of histograms of true (blue) and generated (red) Maxwell-J\"{u}ttner distribution for different values of $\Theta$ parameter.}
\end{figure*}

\begin{figure*}[htb]
    \centering
    \includegraphics[width=.22\linewidth]{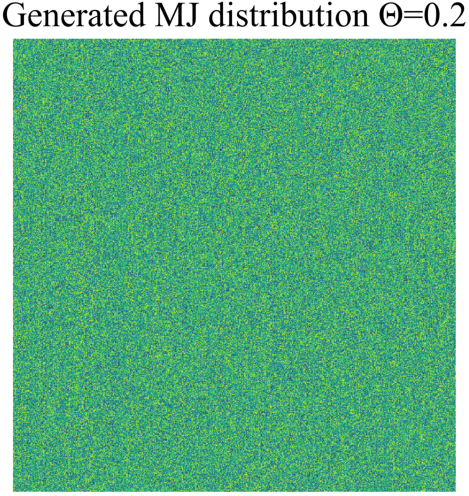}\hfill%
    \includegraphics[width=.22\linewidth]{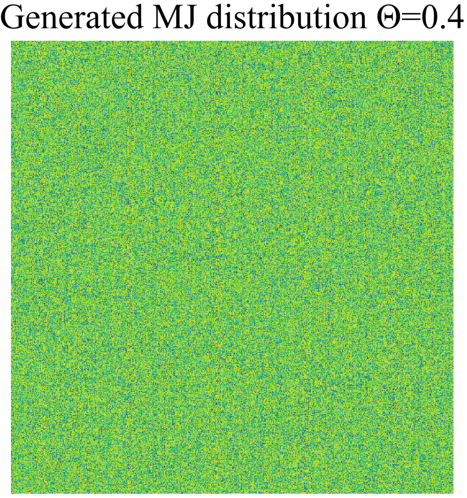}\hfill%
    \includegraphics[width=.22\linewidth]{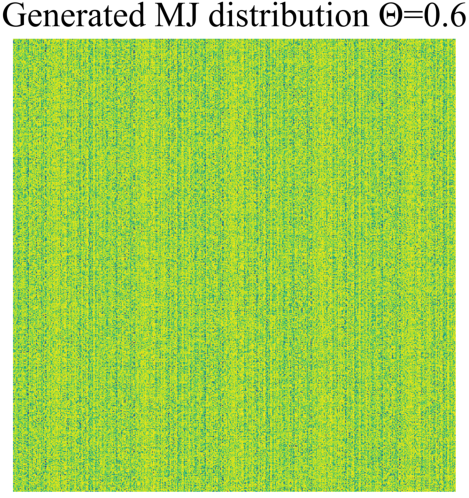}\hfill%
    \includegraphics[width=.22\linewidth]{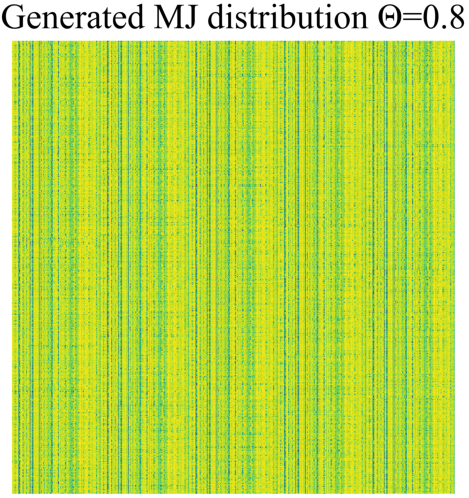}
    \label{fig:vis} 
    \caption{Visual representation of generated samples from Maxwell-J\"{u}ttner distribution with increasing value of $\Theta$. The randomness is good for lower values of $\Delta$, but for higher values of $\Theta$, we can see clear correlations.}
\end{figure*}

The resulting histograms for different values of $\Theta$ are shown in the Figure~\ref{fig:hst}. Histograms show that for all tested values of $\Theta$, the generator is able to produce samples that approximate the MJ distribution well. When individual samples produced by the generator are compared to samples from the MJ distribution using the Kolmogorov-Smirnov (KS) test, we get $P_\mathrm{\Theta=0.2}=0.56$; $P_\mathrm{\Theta=0.4}=0.53$; $P_\mathrm{\Theta=0.6}=0.45$; $P_\mathrm{\Theta=0.8}=0.43$. These KS values show that individual samples generated by the network cannot be distinguished from true distribution. However, when multiple samples are combined, the KS test rejects generated samples confidently. 

Visual depiction of the randomness of the generated samples is shown in Figure~\ref{fig:vis}. For lower values of $\Theta<0.5$ the generated samples do not show any visible pattern or correlation, this is not true for values $\Theta>0.5$ as shown in Fig.~\ref{fig:vis} by two pictures on the right, where there is clear correlation.

The correlation in generated samples, while those samples pass the KS test, suggests a mode collapse of the generator, where the generator identifies one or few solutions with which it is able to fool the discriminator. To mitigate this, we have tried Wasserstein GAN as suggested by \citet{arjovsky2017wasserstein} without success, however.

\section{Conclusions}
We have applied GAN to the problem of sampling the Maxwell-J\"{u}ttner (MJ) distribution that describes electron speed in a relativistic regime. We have shown that a multilayer perception network can learn the MJ distribution even for skewed distribution for highly relativistic electrons. The Kolmogorov-Smirnov test performed on the samples generated by the network cannot distinguish between true and generated samples. However, when the test is performed on a larger set of samples, the test fails.  

In future work, the neural network used should be expanded to include variable parameter $\Theta$ as a hyperparameter. Furthermore, the training process needs to be adjusted to counter mode collapse visible for higher values of parameter $\Theta$.

\bibliography{P418_arxiv}

\begin{thebibliography}{}

\bibitem[Arjovsky et~al., 2017]{arjovsky2017wasserstein}
Arjovsky, M., Chintala, S., and Bottou, L. (2017).
\newblock Wasserstein gan.

\bibitem[{Chakraborty} et~al., 2023]{chakraborty2023years}
{Chakraborty}, T., {Reddy}, U., {Naik}, S., {Panja}, M., and {Manvitha}, B.
  (2023).
\newblock Ten years of generative adversarial nets (gans): A survey of the
  state-of-the-art.

\bibitem[Paganini et~al., 2018]{PhysRevD.97.014021}
Paganini, M., de~Oliveira, L., and Nachman, B. (2018).
\newblock Calogan: Simulating 3d high energy particle showers in multilayer
  electromagnetic calorimeters with generative adversarial networks.
\newblock {\em Phys. Rev. D}, 97:014021.

\bibitem[Schnittman and Krolik, 2013]{Schnittman_2013}
Schnittman, J.~D. and Krolik, J.~H. (2013).
\newblock A monte carlo code for relativistic radiation transport around kerr
  black holes.
\newblock {\em The Astrophysical Journal}, 777(1):11.

\bibitem[Zaheer et~al., 2017]{zaheer2017gan}
Zaheer, M., Li, C.-L., PÃ³czos, B., and Salakhutdinov, R. (2017).
\newblock Gan connoisseur: Can gans learn simple 1d parametric distributions?
\newblock In {\em NIPS Workshop on Deep Learning: Bridging Theory and
  Practice}.

\bibitem[Zenitani, 2015]{10.1063/1.4919383}
Zenitani, S. (2015).
\newblock {Loading relativistic Maxwell distributions in particle simulations}.
\newblock {\em Physics of Plasmas}, 22(4):042116.

\end{thebibliography}
\end{document}